\documentstyle[epsfig,12pt]{article}
\begin{document}
\noindent
ISSN 0284 - 2769 \hfill{TSL/ISV-93-0090}

\hfill{December 1993}
%\maketitle
\vskip1cm
\begin{center}
{\large \bf SHADOWING IN THE DEUTERON\\AND THE NEW $F_2^n/F_2^p$ MEASUREMENTS}
\vskip0.5cm
{\large B. Bade\l{}ek$^1$} and {\large J. Kwieci\'nski$^2$}
\end{center}

\vskip0.5cm
\noindent
$^1$ {\it Department of Pysics, Uppsala University, Uppsala, Sweden}

\noindent
\hspace{2mm}
and {\it Institute of Experimental Physics, Warsaw University, Warsaw,
Poland}\\
$^2$ {\it H. Niewodnicza\'nski Institute of Nuclear Physics, Cracow, Poland}
\\

\medskip\medskip\medskip
\vskip1cm
\begin{abstract}
The quantity $2F_2^d(x,Q^2)/F_2^p(x,Q^2)-1$ is calculated in the
region of low $x$ and low- and moderate $Q^2$ relevant for recent NMC and E665
measurements as well as for the expected final results of the precise NMC
analysis of their low $x$ data.
The calculations include nuclear shadowing
effects and a suitable extrapolation of the structure
functions of free nucleons to the low $Q^2$ region.  The
theoretical results are in a good agreement with the NMC
data. The shadowing correction to the experimental estimate of the
Gottfried sum is quantified.
\end{abstract}

\medskip \medskip \medskip
A new determination of $F_2^p-F_2^n$ at $Q^2=$4 GeV$^2$ by the NMC,
\cite{nmcrapid}, as well as the ongoing NMC data analysis which will result
in extending the precise $F_2^d/F_2^p$ measurements down to $x=$0.0008
\cite{t14} are the main reasons for revisiting the problem of nuclear shadowing
in the deuteron and in particular the way shadowing affects determination
of the Gottfried sum $S_G$:
\begin{equation}
S_G(Q^2)=\int^1_{0}
{dx\over x}\left [F_2^p(x,Q^2) - F_2^n(x,Q^2)\right ]
\label{ig}
\end{equation}
\noindent
where $F_2^p$ and $F_2^n$ are the proton and neutron structure functions.
The NMC performed precise measurements of the ratio $F_2^d/F_2^p$
which permitted to obtain the ratio
$(F_2^n/F_2^p)_{NMC}$ defined as 2$F_2^d/F_2^p$--1,
i.e. no shadowing was assumed.
Taking the shadowing into account the
deuteron structure function $F_2^d(x,Q^2)$ is related in the following
way to the $F_2^p$, $F_2^n$ and to the shadowing term $\delta F_2^d(x,Q^2)$,
which is non-negligible for $x$ less than, say, 0.1 or so:
\begin{equation}
2F_2^d(x,Q^2) = F_2^p(x,Q^2) + F_2^n(x,Q^2) - 2\delta F_2^d(x,Q^2).
\label{ddef}
\end{equation}
Throughout this paper $F_2^d$ and $\delta F_2^d$
will be normalised per nucleon.
It follows from the eq.(\ref{ddef}) that the quantity
$2F_2^d(x,Q^2)/F_2^p(x,Q^2)-1$ which in absence of
nuclear shadowing would be equal to the ratio of the neutron and proton
structure functions, contains also the shadowing effects i.e.:
\begin{equation}
2{F_2^d(x,Q^2)\over F_2^p(x,Q^2)}-1={F_2^n(x,Q^2)\over F_2^p(x,Q^2)}-
{2\delta F_2^d(x,Q^2)\over F_2^p(x,Q^2)}
\label{rationp}
\end{equation}
\noindent
The difference between  the proton and neutron structure functions or the
integrand in the Gottfried sum, eq.(\ref{ig}), is thus the following:
\begin{equation}
F_2^p(x,Q^2)-F_2^n(x,Q^2)=
\left (F_2^p(x,Q^2)-F_2^n(x,Q^2)\right )_{NMC}
- 2\delta F_2^d(x,Q^2)
\label{f2pn}
\end{equation}
\noindent
The $\left (F_2^p(x,Q^2)-F_2^n(x,Q^2)\right )_{NMC}$ term
which in the NMC analyses was obtained as
\begin{equation}
(F_2^p-F_2^n)_{NMC}=2F_2^d{\left[1-{\left (F_2^n\over
F_2^p\right)_{NMC}}\right]
                    \over\left[1+{\left (F_2^n\over F_2^p\right)_{NMC}}\right]}
\label{npexp}
\end{equation}
and which is used by the NMC for determination of the Gottfried sum
has to be corrected for shadowing effects in order to obtain the difference
of the structure functions of free nucleons. Thus
the shadowing leads to {\bf smaller} $S_G(Q^2)$   than that determined
experimentally assuming no shadowing.

\medskip\medskip
In the original NMC determination of $S_G$, \cite{nmcold}, $F_2^d$ in
eq.(\ref{npexp}) was taken
from a global fit to the results of earlier experiments; in the recent
determination, \cite{nmcrapid}, the NMC used the
 $F_2^d$ resulting from a fit to their own precise measurements and to the data
of the SLAC and BCDMS experiments, \cite{nmcf2}. The results of this fit
differed substantially from the parametrisation used in \cite{nmcold},
especially at low $x$.
Also the ratio $F_2^d/F_2^p$ was newly determined in \cite{nmcrapid}:
the statistics was increased and the Dubna scheme instead of the Mo and Tsai
method was used for radiative corrections determination \cite{radcor}.
The measurements of the $F_2^d/F_2^p$ ratio have also been reported recently
by the E665 Collaboration at Fermilab, \cite{e665}. The data reach
$x=$0.00005 at $Q^2=$0.03 GeV$^2$. Precision of these measurements is
however very poor.

\medskip\medskip
The re-evaluated NMC data of ref.\cite{nmcrapid} and the E665 results
\cite{e665} will be compared with our model calculations which include
nuclear shadowing effects in the deuteron and a suitable extrapolation
of the high $Q^2$ structure functions of free nucleons to the low $Q^2$
region. Preditions will also be given for the $(x,Q^2)$ region relevant
for the (expected) final results of the NMC analysis which
will extend their $F_2^d/F_2^p$ measurements down to $x=$0.0008, \cite{t14} and
low values of $Q^2$ ($<Q^2> \approx $0.25 GeV$^2$).

\medskip\medskip
In our model \cite{bbjkshad} the shadowing phenomenon
in the inelastic lepton--deuteron scattering is analysed using the double
interaction
formalism in which we relate shadowing to inclusive diffractive processes.
Both the vector meson and parton contributions are considered for low and
high $Q^2$ values. The QCD corrections with parton recombination are included
for high $Q^2$.

\medskip\medskip
The shadowing term $\delta F_2^d(x,Q^2)$ is related through the (double)
pomeron exchange to the
diffractive structure function, ${\partial^2F_2^{diff} / \partial \xi
\partial t}$
\begin{equation}
\delta F_2^d(x,Q^2) = \int d^2{\bf k_{\bot}} \int^1_{\xi_0} d\xi S(k^2)
{\partial^2F_2^{diff} \over \partial \xi\partial t}
\label{shad}
\end{equation}
where $S(k^2)$ is the deuteron form factor. We define $\xi=2kq/p_dq$ where
$k,q$
and $p_d$ are the four momenta of the pomeron, virtual photon and of the
deuteron respectively; $\xi_0=x(1+M^2_{x0}/Q^2)$ where $M^2_{x0}$ is the lowest
mass squared of the diffractively produced hadronic system. We also have
$t\cong -k^2_{\bot}-k^2_{\parallel}$ where $k^2_{\parallel}=M^2\xi^2$ with $M$
being equal to the nucleon mass.
The integration over $\xi$ corresponds to the integration over $M_x^2$
where $M_x$ is the mass of the diffractively produced system. The region of
low $M_x^2$ is dominated by the diffractive production of low mass vector
mesons
which is assumed to be described by the vector meson dominance mechanism, VMD.
In this model the contribution of the vector mesons to the nuclear shadowing is
\begin{equation}
\delta F_{2v}^d={Q^2\over {4\pi}}\sum_v{M_v^4\delta \sigma_v^d\over \gamma_v^2
(Q^2+M_v^2)^2}
\label{shadv}
\end{equation}
where the sum extends over the low mass vector mesons $\rho, \omega$ and
$\phi$. $M_v$ are the
masses of the vector mesons $v$  and
$\gamma_v$ can be obtained from the leptonic widths of the
vector mesons \cite{bauer}. The double scattering cross section $\delta \sigma
_v^d$ is obtained
from the Glauber model, \cite{glauber} with the energy--dependent vector
meson--nucleon cross sections.
The contribution of large $M_x^2$ corresponds to the partonic mechanism
of shadowing which is related to the partonic content of the pomeron.
For the detailed formulation of the model see references \cite{bbjkshad}.

\medskip\medskip
In \cite{bbjkshad} we have only calculated the
 shadowing term $\delta F_2^d(x,Q^2)$ alone.
In order to make the theoretical prediction for the quantity
$2F_2^d(x,Q^2)/F_2^p(x,Q^2)-1$ and compare it with the existing
data one also needs the
elementary structure functions $F_2^p(x,Q^2)$ and $F_2^n(x,Q^2)$  in particular
in the region of small $x$ and small $Q^2$.  To this aim we shall use the
structure function
model developed by us in \cite{bbjkf2} which satisfactorily reproduces the
structure function data and in which the  contributions from both
the parton model with QCD corrections suitably extended to the low $Q^2$ region
and from the low mass vector mesons were taken into account.
In this model the structure function $F_2(x,Q^2)$ is given by
the following formula:
\begin{eqnarray}
F_2^{p,n}(x=Q^2/(W^2+Q^2-M^2),Q^2)&=& \nonumber \\
{Q^2\over 4\pi}\sum_v{M_v^4\sigma_v(W^2)\over \gamma_v^2
(Q^2+M_v^2)^2}+{Q^2\over (Q^2+Q_0^2)}F_{2AS}^{p,n} (\bar x, Q^2+Q_0^2)&=&
F_{2v}(x,Q^2)+F^{p,n}_{2p}(x,Q^2)
\label{f2mod}
\end{eqnarray}
where the first term corresponds to the VMD contribution
and the second one to the partonic
mechanism. The cross sections $\sigma_v(W^2)$ are vector
meson--nucleon cross sections, $W$ being the effective mass of the
electroproduced system.
The structure functions $F_{2AS}^{p,n} ( x, Q^2)$
are assumed to be given.  They are obtained from the structure
function ana\-ly\-sis in the large $Q^2$ region and are constrained
by the
Altarelli-Parisi evolution equations.  The variable $\bar x$ is
defined as
\begin{equation}
\bar x=x{Q^2+Q_0^2\over Q^2+xQ_0^2}
\label{xbar}
\end{equation}
The parameter $Q_0^2$ was set equal to 1.2 GeV$^2$. It should be observed
that apart from the parameter $Q_0^2$ which is constrained by physical
requirements the representation (\ref{f2mod}) does not contain any other
free parameters except of course those which are implicitly present in the
parametrisation of parton distributions defining $F_{2AS}^{p,n}$.

\medskip\medskip
The shadowing effects in the deuteron have also been considered in refs
\cite{nnzol,Thomas} using the formalism similar to ours, \cite{bbjkshad}.
In \cite{nnzol} the two gluon exchange model of the inelastic diffraction
was used; the VMD contribution to shadowing, corresponding to the double
scattering of vector mesons -- an important part of the shadowing effect at
low $Q^2$  -- was however ignored. This model has recently been used to
estimate
the shadowing effects in the kinematical region of the reanalysed NMC results,
\cite{torino}. In \cite{Thomas} our model
\cite{bbjkshad} was extended by including the meson exchange effects.

\medskip\medskip
In order to do the comparison with the data we computed the values of the
quantity $2F_2^d(x,Q^2)/\-F_2^p(x,Q^2)-1$, eq.(\ref{rationp}),
for the $(x,Q^2=$4 GeV$^2)$ values measured by the NMC,\cite{nmcrapid}. The
$\delta F_2^d$ was calculated using the formulae (\ref{shad}) and
(\ref{shadv}),
\cite{bbjkshad}, and
$F_2^p$ and $F_2^n$ were calculated according to eqs (\ref{f2mod},\ref{xbar}),
 \cite{bbjkf2}.
As a large $Q^2$ structure functions $F_{2AS}^{p,n}$ required by the model
as an input, eq.(\ref{f2mod}), we used the MRS D--' parametrisation
\cite{dmrs},
resulting from the analysis which includes the NMC $F_2$ measurements
\cite{nmcf2} corrected for nuclear shadowing in the large $Q^2$ region.

\medskip\medskip
The results of our calculations are shown in table 1 and in figures 1
and 2. Only $x$ values smaller than about 0.1 where the shadowing is non--
negligible were considered.
 Table 1 contains results corresponding to the re-evaluated NMC data, \cite
{nmcrapid}.
%
%%%%%%%%%%%%%%%%%%%%%%%%%%%%%%%%%%%%%%%%%%%%%%%%%%%%%%%
%
\setcounter{table}{0}
\begin{table}[h]
%\label{tab1}
\begin{center}
\begin{tabular}{|c|c|c|c|c|c|}
\hline
     $x$ & $F_2^p$ & $(F_2^p+F_2^n)/$2 & 2$F_2^d/F_2^p$-1 & $\delta F_2^d$
         & $\Delta S_G$ \\
\hline
\hline
    $0.007$ & $0.4358$  & $0.4315$  & $0.9569$ & $0.0051$ & 0.0264 \\
    $0.015$ & $0.3932$  & $0.3872$  & $0.9477$ & $0.0043$ & 0.0169 \\
    $0.030$ & $0.3712$  & $0.3621$  & $0.9326$ & $0.0035$ & 0.0108 \\
    $0.050$ & $0.3616$  & $0.3479$  & $0.9094$ & $0.0027$ & 0.0059 \\
    $0.080$ & $0.3543$  & $0.3336$  & $0.8714$ & $0.0020$ & 0.0037 \\
    $0.125$ & $0.3436$  & $0.3133$  & $0.8163$ & $0.0012$ & 0.0009 \\
\hline
\hline
\end{tabular}
\caption{\it The proton structure function, $F_2^p$, the structure function
$(F_2^p+F_2^n)/2$,
the quantity $2F_2^d/F_2^p-1$, eq.(\protect\ref{rationp}),
the shadowing contribution to the deuteron structure function
$\delta F_2^d$, \protect\cite{bbjkshad,bbjkf2} and
the shadowing contribution to the Gottfried sum, $\Delta S_G(x_{min} - 0.8) =
2\int_{x_{min}}^{0.8} \delta F_2^ddx/x$ where $x_{min}$ is equal to the lower
limit of the each $x$ bin. The MRS D-' parametrisation \protect\cite{dmrs}
of the high $Q^2$ structure function was used in the calculations.
All values are at $Q^2=$4 GeV$^2$; the $x$ values are the same as for the
reanalysed NMC results \protect\cite{nmcrapid} except that only points
with $x$ smaller than about 0.1 are listed.}
\end{center}
\end{table}
%
%%%%%%%%%%%%%%%%%%%%%%%%%%%%%%%%%%%%%%%%%%%%%%%%%%%%%%%%%%%%%%
%
In fig.1 results of the calculations are presented together with the
NMC \cite{nmcrapid} and E665 \cite{e665} data.
Plotted is  the quantity $2F_2^d(x,Q^2)
/F_2^p(x,Q^2)-1$ which in the absence of shadowing would be just equal to
$F_2^n(x,Q^2)
/F_2^p(x,Q^2)$, cf. eq. (\ref{rationp}).
The average $Q^2$ values for the E665 measurements range from 0.03 GeV$^2$ at
0.00005 to 28.8 GeV$^2$ at $x=$0.15, i.e. the very low $x$ data were taken
at very low $Q^2$. The continuous curve is a result
of our calculations which
were done in the $(x,Q^2=$4GeV$^2)$ points corresponding to the NMC results
and the broken one to the $(x,Q^2)$ points corresponding to the
E665 measurements.

\medskip\medskip
It can be noticed from the table 1 and figure 1 that the quantity
$2F_2^d(x,Q^2)/F_2^p(x,Q^2)-1$ is close to one
in the region of
small $x$ and its difference from unity has to be attributed to the
shadowing effect. The reason is that the difference between elementary
structure functions $F_2^p$ and $F_2^n$ calculated from equations
(\ref{f2mod}) and (\ref{xbar}) is very small in the low $x$, low $Q^2$
region. The function $F_2^p-F_2^n$ should of course vanish
in the limit $x=$0 (for fixed $Q^2$) since it is controlled by the $A_2$
Regge exchange in this limit, i.e.
\begin{equation}
F_2^p-F_2^n \sim x^{1-\alpha_{A_2}}
\end{equation}
where $\alpha_{A_2} \approx$ 1/2 is the $A_2$ reggeon intercept.
The expectations coming from  Regge theory are incorporated in the
parametrisation of parton distributions defining $F_{2AS}^{p,n}$ and
of $\sigma_v(W^2)$.

\medskip\medskip
The model predicts saturation of
shadowing, i.e. its approximate $x$ independence.  This comes
from the fact that the $x$ dependence of the
$\delta F^d_2(x,Q^2)$ and of the elementary structure function
$F_2^p(x,Q^2)$ is similar
%and relatively weak
in the region of $x$ and $Q^2$ relevant for the data.
%This fact is closely related with the
%dominance of the VMD mechanism for both functions in the region
%of low $Q^2$.
The saturation persists to be present for moderately large values of $Q^2$.
%
%%%%%%%%%%%%%%%%%%%%%%%%%%%%%%%%%%%%%%%%%%%%%%%%%%%%%%%%%
%
\begin{figure}[h]
\begin{center}
\mbox{\vspace{1cm} {\epsfxsize=12cm \epsfysize=12cm \epsffile{
badelek_kwiecinski_fig1.eps}}}
\caption{\it The quantity $2F_2^d/F_2^p - 1$ calculated from our models
\protect\cite{bbjkshad,bbjkf2} which include the shadowing in the deuteron,
eq.(\protect\ref{rationp}) compared with the re-evaluated
NMC measurements, \protect\cite{nmcrapid}, (cf. table 1) and with the Fermilab
E665 data, \protect\cite{e665}. The errors are statistical.}
\end{center}
\end{figure}
%
%%%%%%%%%%%%%%%%%%%%%%%%%%%%%%%%%%%%%%%%%%%%%
%
\medskip\medskip
Our models reproduce well the NMC measurements; the E665 results,
reaching down to very low values of $x$ are not sufficiently precise
to conclude about the shadowing. Observe that in the NMC/E665 data overlap
region the theoretical prediction practically coincide. We also made
calculations for the $(x,Q^2)$ points relevant for the extended sample
of the NMC $F_2^d/F_2^p$ measurements, \cite{t14}. The $x$ values reach down to
$x=$0.0008 at $<Q^2>\approx$0.25 GeV$^2$. The $<Q^2>$ there increases
with increasing $x$ up to $<Q^2>\approx$ 10 GeV$^2$ at $x=$0.11. These
predictions were practically indistinguishable from those for the E665, fig.1.

\medskip\medskip
In fig.2 we show the shadowing contribution to the deuteron structure function,
$\delta F_2^d$, calculated according to the model \cite{bbjkshad} in the
$(x,Q^2)$ points corresponding to the extended NMC measurements
(preliminary data \cite{t14}).
At $x\sim$0.001 the shadowing accounts for approximately
2.5$\%$ of the deuteron structure function. Both
the partonic and the vector meson contributions are shown in the figure.
 The main contribution to the
$\delta F_2^d$ comes from the vector meson rescattering, cf. eq.(\ref{shadv});
the parton contributions become comparable only
for $Q^2\sim$2 GeV$^2$ corresponding to  $x\sim$0.01.

\medskip\medskip
The nuclear shadowing effect in the deuteron {\bf lowers} the NMC Gottfried sum
%
%%%%%%%%%%%%%%%%%%%%%%%%%%%%%%%%%%%%%%%%%%%%%%%%%%%%%%%%%%
%
\begin{figure}[ht]
\begin{center}
\mbox{\vspace{1cm} {\epsfxsize=12cm \epsfysize=12cm \epsffile{
badelek_kwiecinski_fig2.eps}}}
\caption{\it The shadowing contribution to the deuteron structure function,
$\delta F_2^d$, calculated according to the model \protect\cite{bbjkshad}
in the $(x,Q^2)$ points corresponding to the extended NMC measurements,
\protect\cite{t14}. The partonic and the vector meson scattering contributions
are also shown.}
\end{center}
\end{figure}
%
%%%%%%%%%%%%%%%%%%%%%%%%%%%%%%%%%%%%%%%%%%%%%%%%%%%%%%%%%%%
%
estimate, cf. eqs (\ref{f2pn}) and (\ref{ig}). The numerical results of the
shadowing correction to the Gottfried sum,
\begin{equation}
\Delta S_G(x_{min} - 0.8) =
2\int_{x_{min}}^{0.8} {dx \over x}\delta F_2^d
\label{delta}
\end{equation}
where $x_{min}$ is equal to the lower limit of the measured $x$ interval,
are given in table 1. This implies that the Gottfried sum, estimated by the NMC
 \cite{nmcrapid} to be equal to $S_G ($0.004 -- 0.8$) =$0.221 $\pm$ 0.008
(stat.)$\pm$ 0.019 (syst.) is thus further {\bf decreased} by
$\Delta S_G=$0.0264, i.e. by about 12$\%$. For a different
estimate of the shadowing effect in the Gottfried sum, \cite{zoller}, the
relevant number is about 17$\%$.

\medskip\medskip
To sum up we have calculated the quantity $2F_2^d(x,Q^2)/F_2^p(x,Q^2)-1$
in the region of low $x$ and low- and moderate $Q^2$ relevant  for the NMC and
E665 measurements and for the expected results of the final NMC analysis of
their low $x$ data.
In this region both the shadowing as well as the elementary structure functions
are dominated by the VMD mechanism. The model predicts saturation, i.e. $x$
independence of  the quantity $2F_2^d(x,Q^2)/F_2^p(x,Q^2)-1$ in the very
low $x$ and $Q^2$ region. The fact that this quantity stays
systematically below unity has to be atributed to nuclear
shadowing in the deuteron.  The amount of shadowing
predicted from the model \cite{bbjkshad,bbjkf2}
is in agreement with the recently re-evaluated NMC data at $Q^2=$4 GeV$^2$.
The shadowing decreases the
experimentally estimated value of the Gottfried sum by about 12$\%$.

\medskip\medskip\medskip\medskip
\noindent
{\Large {\bf Acknowledgement}} \\

\medskip
This research has been supported in part by the Polish Committee
for Scientific Research, grant number 2 P302 062 04.

\end{document}